\documentclass[aps,pre,twocolumn,longbibliography,showpacs,amsmath,amsmath,amssymb,amsfonts,superscriptaddress]{revtex4-1}
\usepackage{graphicx}
\usepackage{dcolumn}
\usepackage{bm}
\usepackage{color}
\usepackage{multirow}
\usepackage{hyperref}

\begin{document}
\title{Physical swap dynamics, shortcuts to relaxation, and entropy production in dissipative Rydberg gases}

\author{Ricardo Guti\'errez}
\affiliation{Complex Systems Group \& GISC, Universidad Rey Juan Carlos, 28933 M\'{o}stoles, Madrid, Spain}
\author{Juan P. Garrahan}
\affiliation{School of Physics and Astronomy, University of Nottingham, Nottingham, NG7 2RD, United Kingdom}
\affiliation{Centre for the Mathematics and Theoretical Physics of Quantum Non-equilibrium Systems, University of Nottingham, Nottingham NG7 2RD, UK}
\author{Igor Lesanovsky}
\affiliation{School of Physics and Astronomy, University of Nottingham, Nottingham, NG7 2RD, United Kingdom}
\affiliation{Centre for the Mathematics and Theoretical Physics of Quantum Non-equilibrium Systems, University of Nottingham, Nottingham NG7 2RD, UK}

\keywords{}
\begin{abstract}
Dense Rydberg gases are out-of-equilibrium systems where strong density-density interactions give rise to effective kinetic constraints. They cause dynamic arrest associated with highly-constrained many-body configurations, leading to slow relaxation and glassy behavior. Multi-component Rydberg gases feature additional long-range interactions such as excitation-exchange. These are analogous to particle swaps used to artificially accelerate relaxation in simulations of atomistic models of classical glass formers. In Rydberg gases, however, swaps are real physical processes, which provide dynamical shortcuts to relaxation. They permit the accelerated approach to stationarity in experiment and at the same time have an impact on the non-equilibrium stationary state. In particular their interplay with radiative decay processes amplifies irreversibility of the dynamics, an effect which we quantify via the entropy production at stationarity. Our work highlights an intriguing analogy between real dynamical processes in Rydberg gases and artificial dynamics underlying advanced Monte Carlo methods. Moreover, it delivers a quantitative characterization of the dramatic effect swaps have on the structure and dynamics of their stationary state.
\end{abstract}


\maketitle

\section{Introduction}

Glasses are out of equilibrium systems typically produced by lowering the temperature of a liquid until the relaxation time exceeds experimentally accessible timescales \cite{binder2011,berthier2011,biroli2013}. Close to the glass transition, the physics is different from that of a high-temperature fluid, displaying dynamic heterogeneity (i.e. the coexistence of space-time regions with remarkably different timescales) arising from effective kinetic constraints \cite{ritort2003,chandler2010}. What makes the dynamics interesting, however, renders the simulation of glassy fluids challenging, as the computational equilibration times also increase dramatically. A method to reduce these times in Monte Carlo simulations is to artificially introduce swaps between particles of different species. This approach was proposed in Ref. \cite{grigera2001}, and later refined in e.g. Refs. \cite{brumer2004,fernandez2007}. It enables the numerical equilibration of strongly arrested systems \cite{ninarello2017}, providing insight into lengthscales characterizing glassiness at low temperatures \cite{biroli2008, gutierrez2015}, the jamming transition at high densities \cite{berthier2016} or the configurational entropy of deeply supercooled states \cite{berthier2017}.

In the domain of cold atomic gases Rydberg atoms have emerged as a platform \cite{gallagher2005,saffman2010,low2012} for studying collective dynamical behaviors \cite{urvoy2015,valado2016,gutierrez2017,schempp2014,gutierrez2016,letscher2017} which are closely linked to that of glasses \cite{lesanovsky2013, lesanovsky2014, gutierrez2015b}. The physics of interacting Rydberg atoms is governed by blockade effects \cite{jaksch2000,lukin2001} analogous to the excluded volume interactions underlying the dynamics of glass-forming liquids and jammed systems of hard objects \cite{binder2011, berthier2011,sanders2015}. Also, in both Rydberg gases \cite{perezespigares2018} and  glassy soft-matter models \cite{garrahan2007, garrahan2009, Hedges2009,Speck2012} dynamical phase transitions have been shown to be at the heart of the observed collective dynamical effects.

In this paper we investigate the impact of excitation swaps on the relaxation timescales and on the structure and dynamics of the stationary state of Rydberg gases. Swaps are caused by resonant dipole-dipole interactions \cite{saffman2010,delesleleuc2017} which effectuate state exchanges over long distances. This is strongly reminiscent of the above-mentioned advanced Monte Carlo approaches in which particle swaps accelerate the approach of highly arrested systems to stationarity. In Rydberg gases swaps are physical and their strength and range can be controlled. We show that their inclusion indeed shortens the characteristic timescales by several orders of magnitude, which allows one to experimentally introduce shortcuts to relaxation. We also show that swaps drive the system further away from equilibrium conditions, in particular in conjunction with radiative decay processes. The resulting irreversibility of the dynamics is quantified via the entropy production. We outline how this quantity may be experimentally accessed and deduce the existence of an irreversibility bound for large swap rate.

\section{Multi-component Rydberg gases}
\subsection{Description of the model}

We consider a system of $N$ atoms. The ground state $\left|0\right>$ of each atom is resonantly coupled by a laser field to the Rydberg states $\left|1\right>, \left|2\right>, \ldots, \left|p\right>$
(with energies $E_0 < E_1 < \cdots < E_p$). Such systems are called multi-component Rydberg gases, as they include different kinds of Rydberg excitations (i.e. atoms excited to  different Rydberg levels), see Fig \ref{figswaps} (a) for an illustration of the $p=2$ case. The Hamiltonian $H = H_0 + H_1$ includes the density-density interactions in $H_0$ and the exchange interactions and the driving in $H_1$. Atoms in the Rydberg states  $|s\rangle$ and  $|s^\prime\rangle$ at positions ${\bf r}_k$ and ${\bf r}_l$ interact through a power-law potential $V_{kl}^{ss^\prime} = C_\alpha^{ss^\prime}/|{\bf r}_k - {\bf r}_l|^\alpha$ with exponent $\alpha$. For simplicity we denote the intra-level interactions by $V_{kl}^{s}$ instead of  $V_{kl}^{ss}$. We thus have 
\begin{equation}
H_0 =  \sum_{k < l} \sum_{s=1}^p \left[ V_{kl}^{s}\, n_s^{(k)} n_s^{(l)} + \sum_{s^\prime \neq s} V^{s s^\prime}_{kl} n_s^{(k)} n_{s^\prime}^{(l)}\right]\!,
\label{h0}
\end{equation}
where $n_s^{(k)}=\left|s\right>_k\!\left<s\right|$, which is diagonal in the eigenbasis $\left|0 0 \ldots 0\right>$, $\left|0 0 \ldots 1\right>$, etc. The off-diagonal part is 
\begin{equation}
H_1 = \sum_{k=1}^N  \sum_{s=1}^p \Omega_s\, \sigma_{sx}^{(k)} + \sum_{k < l}\sum_{s<s^\prime}  E_{kl}^{ss^\prime} \left( \sigma_{ss^\prime}^{(k)} \sigma_{s^\prime s}^{(l)}+ \sigma_{s^\prime s}^{(k)} \sigma_{s s^\prime}^{(l)}\right).
\label{h1}
\end{equation}
Here, $\sigma_{sx}^{(k)} = |0\rangle_k\langle s|+|s\rangle_k\langle 0|$ is the Pauli matrix for a given transition and  $\sigma_{ss^\prime}^{(k)} =  |s\rangle_k\langle s^\prime |$. The first term in Eq.~(\ref{h1}) includes the driving in the rotating wave approximation,  where $\Omega_s$ is the Rabi frequency of the coupling between $\left|s\right>$ and $\left|0\right>$.  The second term contains the exchange interactions, with a power-law dependence on the distance between atoms $E_{kl}^{ss^\prime} =  \mathcal{E}_\beta^{ss^\prime}/|{\bf r}_k - {\bf r}_l|^\beta$, with exponent $\beta$.

The full dynamics is governed by a quantum master equation in Lindblad form, which includes as incoherent processes a dephasing term (arising from the laser linewidth, thermal effects, etc. \cite{urvoy2015,valado2016,gutierrez2017}) and a spontaneous decay term for each Rydberg level. This master equation can be written as $\partial_t \rho = \mathcal{L} \rho = \mathcal{L}_0 \rho + \mathcal{L}_1 \rho$. Here, the diagonal Liouvillian $\mathcal{L}_0$ contains not only the interaction Hamiltonian $H_0$ but also a dissipative term that governs the dephasing process, and is defined as
\begin{equation}
\mathcal{L}_0 \rho = -i[H_0,\rho] + \sum_{s=1}^p  \gamma_s \sum_{k=1}^N\left(n_s^{(k)} \rho\, n_s^{(k)} - \frac{1}{2} \left\{n_s^{(k)},\rho\right\}\right)\!,
\label{l0pspec}
\end{equation}
where $\gamma_s$ is the dephasing rate of $\left|s\right>$ w.r.t. $\left|0\right>$. 
The superoperator $\mathcal{L}_0$ therefore consists of a Hamiltonian part and a dissipator whose individual terms commute. On the other hand, the Liouvillian $\mathcal{L}_1$ includes the evolution due to the off-diagonal part of the Hamiltonian $H_1$ and a dissipator that accounts for spontaneous decay in the system. It takes the form
\begin{equation}
\mathcal{L}_1 \rho = -i[H_1,\rho] + \sum_{s=1}^p  \kappa_s \sum_{k=1}^N\left(\sigma_{-s}^{(k)} \rho\, \sigma_{+s}^{(k)} - \frac{1}{2} \left\{n_s^{(k)},\rho\right\}\right),
\label{l0pspec}
\end{equation}
where $\sigma_{-s}^{(k)} = |0\rangle_k\langle s|$, $\sigma_{+s}^{(k)} = {\sigma_{-s}^{(k)}}^{\dag}$, and $\kappa_s$ is the decay rate of excited level $\left|s\right>$.

\subsection{Perturbative analysis based on projection operators}

When different dynamical processes evolve on vastly different timescales, one sometimes can  adiabatically eliminate fast-evolving degrees of freedom, and focus on the evolution of the remaining (slow) ones. In this section, we derive the effective dynamics in the limit of strong dissipation and strong (density-density) interactions, where the non-diagonal terms of the full Liouvillian $\mathcal{L}$, which are contained in $\mathcal{L}_1$ can be treated perturbatively: $\Omega_s, \kappa_s, \mathcal{E}_\beta^{s s^\prime} \ll \gamma_s, C_\alpha^{s s^\prime}$. In this strongly-dissipative limit, coherent superpositions of local atomic states dephase exponentially fast, allowing for a description of the dynamics that only includes (slow-evolving) diagonal elements of the density matrix on timescales larger than $\gamma_s^{-1}$ (for the atomic level with the smallest dephasing rate $\gamma_s$ in the system). An analogous approach has been previously used to explore multi-component  Rydberg gases in the absence of exchange processes and decay \cite{gutierrez2016}, and also to study one-component systems \cite{lesanovsky2013,cai2013,petrosyan2013,marcuzzi2014,schonleber2014,hoening2014}.  

We start by working out the effect of dephasing on the dynamics.  Using the notation, $\mathcal{L}_{0,d}^k = \sum_{s=1}^p \left[ \gamma_s \left(n_s^{(k)} \rho\, n_s^{(k)} - \frac{1}{2} \left\{n_s^{(k)},\rho\right\}\right)\right]$, the action of the Liouvillian $\mathcal{L}_0$ leads to the evolution $e^{\mathcal{L}_0 t} \rho = e^{-i H_0 t}\, [ \bigotimes_k e^{\mathcal{L}_{0,d}^k t} \rho]\, e^{i H_0 t}$
where the term in brackets gives
\begin{equation}
\bigotimes_{l \neq k} e^{\mathcal{L}_{0,d}^l t}\! 
\begingroup 
\setlength\arraycolsep{0pt}
\begin{pmatrix}
\rho_{pp}^{(k)} & e^{-\frac{(\gamma_{p-1}\!+\!\gamma_p) t}{2}} \rho_{p(p-1)}^{(k)}  & \cdots  \\
e^{-\frac{(\gamma_{p-1}+\gamma_p) t}{2}} \rho_{(p-1)p}^{(k)} & \rho_{(p-1)(p-1)}^{(k)} & \cdots \\
\vdots & \vdots & \ddots  \end{pmatrix}\endgroup.
\label{actiondissippspec}
\end{equation}
Here, $\rho_{mn}^{(k)}$ are the $p^{N-1} \times p^{N-1}$ matrices defined by $\rho_{mn}^{(k)} = {}_k{\left< n\right|} \rho \left|m\right>_k$, using as basis states $\left|0\right>_k$, $\left|1\right>_k$, $\left|2\right>_k$, $\ldots,$ $\left|p\right>_k$. The off-diagonal entries of the density matrix are seen to decay exponentially, a fact that is not altered by the action of the coherent dynamics $e^{-i H_0 t}$. Therefore, the evolution under $\mathcal{L}_0$ becomes, at times considerably larger than the inverse of the dephasing rates, a projector $\mathcal{P}$ on the diagonal of $\rho$, $\mathcal{P}\rho = \lim_{t\to\infty} e^{\mathcal{L}_0 t} \rho = \textrm{diag}(\rho)$, as happens in the case of just one Rydberg level \cite{lesanovsky2013}. The removal of all coherences leads to a diagonal density matrix, where each classically accessible configuration (e.g. $\left|0 0 1 0 2 0 3 \cdots\right>$) is given a certain probability of occurrence.

By using the operator $\mathcal{P}$ (which projects on the slow-evolving diagonal) and its complement $\mathcal{Q} = 1-\mathcal{P}$ (which projects on the rest of the Liouville space, where rapidly-decaying quantum coherences reside), we can formulate the effective evolution equation for the diagonal density matrix $\mu = \mathcal{P}\rho$, which describes the slow evolution of the excitation dynamics. To second order in $\mathcal{L}_1$, the general Nakajima-Zwanzig expression \cite{breuer2002} is given by
\begin{equation}
\partial_t \mu = \mathcal{P}\mathcal{L}_1\mu + \int_0^\infty dt\, \mathcal{P} \mathcal{L}_1 \mathcal{Q} e^{\mathcal{L}_0 t} \mathcal{Q} \mathcal{L}_1 \mu. 
\label{effectivedynpre}
\end{equation}
See \cite{marcuzzi2014} for a detailed treatment of the perturbative expansion in the context of Rydberg gases. In the present case, the first term simplifies to
\begin{equation}
\mathcal{P}\mathcal{L}_1\mu = \sum_{s=1}^p \kappa_s \sum_{k=1}^N \left(\sigma_{-s}^{(k)}\mu\, \sigma_{+s}^{(k)} - n_s^{(k)} \mu\right).
\label{decproc}
\end{equation}
As $H_1$ takes the elements of $\mu$ off the main diagonal, its action vanishes under the projection $\mathcal{P}$. On the other hand, the integrand in Eq.~(\ref{effectivedynpre}) reduces to
\begin{widetext}
\begin{equation}
\mathcal{P} \mathcal{L}_1 \mathcal{Q} e^{\mathcal{L}_0 t} \mathcal{Q} \mathcal{L}_1 \mu = \underbrace{-  \sum_{s}  \sum_{k}  \Omega_{s}^2\,  \mathcal{P} [\sigma_{s x}^{(k)}\,, e^{\mathcal{L}_0 t}  [\sigma_{s x}^{(k)}, \mu]]}_{D[\mu,t]}  \underbrace{- \sum_{s<s^\prime} \sum_{k < l}\left(E_{kl}^{ss^\prime}\right)^2 \mathcal{P} [\sigma_{s s^{\prime}}^{(k)} \sigma_{s^{\prime}s}^{(l)}+ \sigma_{s^{\prime}s}^{(k)} \sigma_{s s^{\prime}}^{(l)}\,, e^{\mathcal{L}_0 t}  [\sigma_{ss^\prime}^{(k)} \sigma_{s^\prime s}^{(l)}+ \sigma_{s^\prime s}^{(k)} \sigma_{s s^\prime}^{(l)} , \mu]].}_{E[\mu,t]}
\label{integrand2}
\end{equation}
\end{widetext}
A detailed derivation can be found in Appendix A. For simplicity the first term in Eq.~(\ref{integrand2}), arising from the driving, is denoted as $D[\mu,t]$, and the second, related to exchange processes, as $E[\mu,t]$. Additionally, there is the term that accounts for the decay processes, Eq.~(\ref{decproc}). Putting it all together, Eq.~(\ref{effectivedynpre}) can be rewritten as 
\begin{eqnarray}
\partial_t \mu &=& \int_0^\infty dt\, D[\mu,t] +\int_0^\infty dt\, E[\mu,t]\nonumber\\
&+& \sum_{s=1}^p \kappa_s \sum_{k=1}^N \left(\sigma_{-s}^{(k)}\mu\, \sigma_{+s}^{(k)} - n_s^{(k)} \mu\right).
\label{effectivedynpre2}
\end{eqnarray}

\subsection{Effective dynamics: driving}

The laser driving term $\int_0^\infty dt\, D[\mu,t]$ in Eq.~(\ref{effectivedynpre2}), gives rise to the excitations and de-excitations in the system. While this has already been calculated before \cite{gutierrez2016}, we include the details in Appendix B for completeness. It yields a classical stochastic generator
\begin{equation}
\int_0^\infty dt\, D[\mu,t] = \sum_{s=1}^p \frac{4\Omega_s^2}{\gamma_s} \sum_k \Gamma_{s}^{(k)} \left[\sigma_{sx}^{(k)} \mu \sigma_{sx}^{(k)} -  \mathcal{I}_s^{(k)} \mu \right]\!,
\label{effectiveintpspecD}
\end{equation}
where the projection operator $\mathcal{I}_s^{(k)} = n_{s}^{(k)} + |0\rangle_k \langle 0|$ cancels all the elements in $\mu$ that do not correspond to the ground state or $|s\rangle$ at site $k$. The rates for a transition $|0\rangle \to |s\rangle$ or $|s\rangle \to |0\rangle$ at site $k$ are
\begin{equation}
\Gamma_{s}^{(k)} = \frac{1}{1 + \left(\frac{2 \mathcal{V}_s^k}{\gamma_s}\right)^2}.
\label{preratesintpspec}
\end{equation}
The increase in the interaction energy that is required for the excitation of atom $k$ to level $|s\rangle$ has been denoted as  $\mathcal{V}^k_s = \sum_m \left[ V^s_{km} n_s^{(m)} + \sum_{s^\prime\neq s} V^{ss^\prime}_{km} n_{s^\prime}^{(m)}\right]$ for convenience.

\subsection{Effective dynamics: exchange}

The term $\int_0^\infty\! dt\, E[\mu,t]$ in the effective dynamics, Eq.~(\ref{effectivedynpre2}), which is due to the exchange interactions in $H_1$, is analyzed in Appendix C. It can also be written as classical stochastic generator
\begin{eqnarray}
&\int_0^\infty\! dt\, E[\mu,t] = \displaystyle\sum_{s<s^\prime} \frac{2 ( \mathcal{E}_\beta^{ss^\prime})^2}{\gamma_s + \gamma_s^\prime} \sum_{k < l} \frac{1}{|{\bf r}_k - {\bf r}_l|^{2\beta}} \Gamma_{ss^\prime}^{(kl)} \times\nonumber\\
& \times\! \left(\!\sigma_{s s^\prime}^{(k)}\! \sigma_{s^\prime s}^{(l)}\, \mu\, \sigma_{s^\prime s}^{(k)} \sigma_{s s^\prime}^{(l)}\! +\! \sigma_{s^\prime s}^{(k)}\! \sigma_{s s^\prime}^{(l)}  \mu\, \sigma_{s s^\prime}^{(k)}\! \sigma_{s^\prime s}^{(l)}\! -\!  \mathcal{I}_{ss^\prime}^{(k,l)} \mu\right)
\label{effectiveintpspecE}
\end{eqnarray}
where $\mathcal{I}_{ss^\prime}^{(k,l)} = n_s^{(k)} n_{s^\prime}^{(l)} +  n_{s^\prime}^{(k)} n_s^{(l)}$. The rate for an exchange $s \leftrightarrow s^\prime$ between sites $k$ and $l$, i.e. the excitation-swap rate, is 
\begin{equation}
\displaystyle\Upsilon_{ss^\prime}^{(kl)} = \frac{2 ( \mathcal{E}_\beta^{ss^\prime})^2}{\gamma_s + \gamma_s^\prime}\frac{1}{|{\bf r}_k - {\bf r}_l|^{2\beta}}\, \Gamma_{ss^\prime}^{(kl)}
\label{ratesE}
\end{equation}
with $\Gamma_{ss^\prime}^{(kl)} = \left({1+\left[\frac{1}{\gamma_s\!+\!\gamma_s^\prime} \left(\mathcal{V}^k_{s^\prime}\!-\!\mathcal{V}^k_s\! +\! \mathcal{V}^l_s\! -\!  \mathcal{V}^l_{s^\prime}\right)\right]^2}\right)^{-1}\!.$

\begin{figure*}[t]
\includegraphics[scale=0.32]{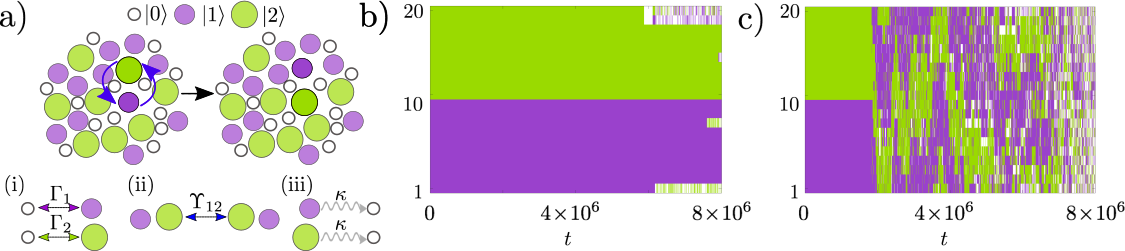}
\vspace{0cm}
\caption{ {\sf \bf Impact of excitation swaps on the dynamics of Rydberg gases.}
(a) An excitation swap in a 2D Rydberg gas (above), and list of relevant processes (below): (i) (de-)excitations, (ii) swaps, and (iii) decay. (b) Representative trajectory of a chain of $N=20$ atoms for $R=4$ and $R_c = 1$ without exchange interactions, $U=0$ (c) Representative trajectory of a chain of $N=20$ atoms for $R=4$ and $R_c = 1$ with exchange interactions, $U=1$. Colors indicate the state of atoms following the legend in (a). In both (b) and (c) decay processes have been excluded, $\kappa = 0$.}\label{figswaps}
\end{figure*}

\subsection{Transition rates}

We conclude that the equation governing the effective dynamics, Eq.~(\ref{effectivedynpre2}), is a classical master equation with three different processes (excitations, swaps and decay) occuring with different rates, see Fig \ref{figswaps} (a). Specifically as follows.

\noindent (i) Excitations and de-excitations between $|0\rangle$ and $|s\rangle$, which, according to Eqs.~(\ref{effectiveintpspecD}, \ref{preratesintpspec}), occur with an inverse timescale $\frac{4\Omega_s^2}{\gamma_s}$, and with a configuration-dependent rate 
\begin{equation}
\frac{1}{\Gamma_{s}^{(k)}}  = 1 +\left[\frac{2}{\gamma_{s} a^\alpha} \sum_{m\neq k} \left(\frac{C_\alpha^{s} n_{s}^{(m)} + \sum_{s^\prime \neq s} C_\alpha^{ss^\prime} n_{s^\prime}^{(m)}}{|\hat{{\bf r}}_k - \hat{{\bf r}}_m|^\alpha} \right)\right]^2.
\label{trans1}
\end{equation}
We have made explicit the power-law dependence on the distance between atoms by giving the positions in a lattice in terms of reduced position vectors $\hat{\bf r}_k = {\bf r}_k/a$, where $a$ is the lattice constant.

\noindent (ii) Excitation swaps $|s\rangle_k |s^\prime\rangle_l \leftrightarrow |s^\prime\rangle_k |s\rangle_l$ (\ref{effectiveintpspecE}). The power-law suppression $\frac{1}{|{\bf r}_k - {\bf r}_l|^{2\beta}}$ in Eq.~(\ref{effectiveintpspecE}) severely limits the possibility of swaps between highly distant atoms. The configuration-dependent part of the swap rates in (\ref{ratesE}), $\Gamma_{ss^\prime}^{(kl)}$, contains in the denominator the difference between the energy associated with the presence of an excitation $|s\rangle$ at site $k$ and an excitation $|s^\prime\rangle$ at site $l$ and that resulting from an exchange of the excited levels of $k$ and $l$ (i.e. the energy before and after a swap). If we make explicit the interactions, the rates appear as
\begin{eqnarray}
&\Upsilon_{ss^\prime}^{(kl)}\! =\! \frac{\frac{2 ( \mathcal{E}_\beta^{ss^\prime})^2}{\gamma_s + \gamma_s^\prime}}{|{\bf r}_k - {\bf r}_l|^{2\beta}}\left\{1\! +\!\left[\! \frac{a^{-\alpha}}{\gamma_s\!+\!\gamma_s^\prime}\! \displaystyle\sum_{m\neq k,l}\! \left(\frac{1}{|\hat{{\bf r}}_k\! -\! \hat{{\bf r}}_m|^\alpha}\! -\! \frac{1}{|\hat{{\bf r}}_l\! -\! \hat{{\bf r}}_m|^\alpha}\right)\right.\right.\!\nonumber\\
&\!\left.\left.\!\left(\! C_\alpha^{s^\prime}\! n_{s^\prime}^{(m)}\!+\!\displaystyle\sum_{s^{\prime\prime}\!\neq s^\prime}\! C_\alpha^{s^\prime\! s^{\prime\prime}}\! n_{s^{\prime\prime}}^{(m)}\!-\! C_\alpha^{s} n_{s}^{(m)}\!-\!\displaystyle\sum_{s^{\prime\prime}\!\neq s}\! C_\alpha^{s s^{\prime\prime}}\! n_{s^{\prime\prime}}^{(m)}\!\right)\! \right]^2\!\right\}^{-1}.
\label{trans2}
\end{eqnarray}

\noindent (iii) Decay processes leading from a given excited level $|s\rangle$ to the ground state $|0\rangle$ with a constant rate $\kappa_s$, which are accounted for by the last term in Eq.~(\ref{effectivedynpre2}).

\subsection{Simplifying parameter choices}

The transition rates in Eqs.~(\ref{trans1}, \ref{trans2}) contain many different parameters associated to processes and interactions involving different Rydberg levels. Due to the high dimensionality of the parameter space, we need to make some simplifying parameter choices in order to explore the dynamics that emerges from excitation, swap and decay processes in numerical simulations. The first simplifying assumption that we make is that all transitions are driven with the same Rabi frequency $\Omega_1 = \Omega_2 = \cdots  = \Omega_p\equiv \Omega$ and dephase with the same rate $\gamma_{1} = \gamma_{2} = \cdots = \gamma_{p} \equiv \gamma$. Moreover, the decay rates are considered to be equal too, $\kappa_{1} = \kappa_{2} = \cdots = \kappa_p \equiv \kappa$. Following Ref.~\cite{gutierrez2016}, we also define an intra-level interaction parameter $R_{s}=a^{-1}[2C^{s}_\alpha/\gamma]^{1/\alpha}$ (for interactions between atoms in the same excited level $s$), and an inter-level interaction parameter $R_{s s^\prime}=a^{-1}[2C^{s s^{\prime}}_\alpha/\gamma]^{1/\alpha}$ (for interactions between atoms in different levels, $s$ and $s^\prime$). These are microscopic (reduced) lengthscales (normalized by the lattice constant) that parametrize the effective range of intra-level and inter-level interactions, respectively, and they determine the dynamical regime of the system \cite{perezespigares2018}.
Moreover, we focus on van der Waals interactions, so the exponents in (\ref{trans1}, \ref{trans2}) satisfy $\alpha = \beta = 6$.


\begin{figure*}[t]
\includegraphics[scale=0.32]{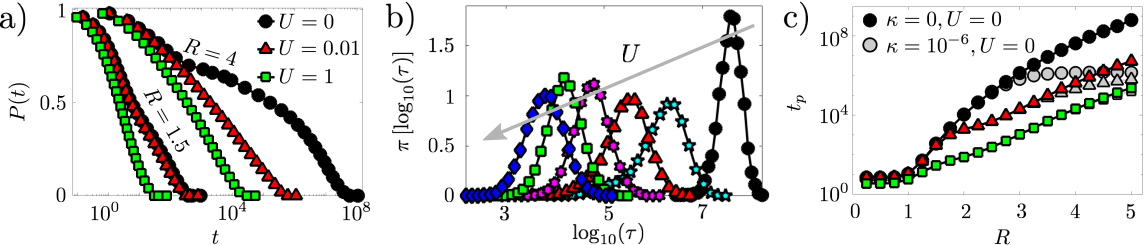}
\vspace{0cm}
\caption{ {\sf \bf Accelerating effect of swaps quantified by the persistence function.}
 (a)  Persistence as a function of time for different values of $U$ (see legend) with $R=1.5$ (left) and $R=4$ (right). (b) Distribution $\pi[\log_{10}(\tau)]$ of local persistence times $\tau$ in logarithmic scale for $U=0$ (black circles), $0.001$ (cyan stars), $0.01$ (red triangles), $0.1$ (magenta hexagons), $1$ (green squares) and $10$ (blue diamonds) (the arrow shows the direction of increasing $U$). (c) Persistence time as a function of $R$ for different values of $U$. Colored symbols [as in legend of panel (a)]  correspond to $\kappa = 0$ (without decay), while gray symbols correspond to a decay rate $\kappa = 10^{-6}$. All panels show numerical results based on a chain of $N=50$ atoms. 200 realizations starting from random initial conditions have been averaged in (a) and (c), while at least $5000$ are used for the histograms in  (b).}\label{figpers}
\end{figure*}

\subsection{Full equations of motion for a two-component Rydberg gas}

For simplicity, our numerical results focus on the case of two components (i.e. two Rydberg levels), $p=2$, with a single intra-level interaction parameter $R$ ($R_1 = R_2 \equiv R$, $C_6^1 = C_6^2$), and an inter-level interaction parameter which we denote by $R_c$ ($R_{12} \equiv R_c$) for convenience. A simpler version of this system without excitation swaps or decay was explored in Ref. \cite{gutierrez2016}. 

The effective dynamics is governed by the following classical master equation
\begin{eqnarray}
\partial_t \mu &=& \frac{4\Omega^2}{\gamma} \sum_k \Gamma_{1}^{(k)} \left[\sigma_{1x}^{(k)} \mu \sigma_{1x}^{(k)} -  \mathcal{I}_1^{(k)} \mu \right]\nonumber\\
&+&\frac{4\Omega^2}{\gamma} \sum_k \Gamma_{2}^{(k)} \left[\sigma_{2x}^{(k)} \mu \sigma_{2x}^{(k)} -  \mathcal{I}_2^{(k)} \mu \right]\nonumber\\ 
&+&\! \sum_{k < l}\! \Upsilon_{12}^{(kl)}\!  \left(\sigma_{12}^{(k)} \sigma_{21}^{(l)} \mu \sigma_{21}^{(k)} \sigma_{12}^{(l)}\! +\! \sigma_{21}^{(k)} \sigma_{12}^{(l)}  \mu \sigma_{12}^{(k)} \sigma_{21}^{(l)}\!-\!  \mathcal{I}_{12}^{(k,l)}\! \mu\right)\nonumber\\ 
&+&  \kappa \sum_{k=1}^N \left(\sigma_{-1}^{(k)}\mu\, \sigma_{+1}^{(k)} - n_1^{(k)} \mu\right)\nonumber\\
&+& \kappa \sum_{k=1}^N \left(\sigma_{-2}^{(k)}\mu\, \sigma_{+2}^{(k)} - n_2^{(k)} \mu\right).
\label{effective2comp}
\end{eqnarray}
The time variable can be rescaled by $4\Omega^2/\gamma$, which is the common factor appearing in all excitation and de-excitation processes (i.e. the first two terms above). Then the factor $(\mathcal{E}_\beta^{12})^2/\gamma$ in Eq.~(\ref{trans2}) is rescaled to $(\mathcal{E}_\beta^{12})^2/4\Omega^2 \equiv U$, which we take as the overall swap strength.

The resulting Markovian dynamics proceeds along classical (diagonal) states,  e.g. $|0 1 2 1 0 0 \cdots \rangle$, and includes the following processes [see Fig.~\ref{figswaps} (a)]: (i) driven (de-)excitations ($|0\rangle \leftrightarrow |1\rangle$, $|0\rangle \leftrightarrow |2\rangle$) with rates [see (\ref{trans1})]
\begin{equation}
\Gamma_{1,2}^{(k)} =  \left(1 + \left[\sum_{m\neq k} \frac{R^6 n_{1,2}^{(m)}+ R_c^6 n_{2,1}^{(m)}}{|\hat{\bf r}_k - \hat{\bf r}_m|^6}\right]^2\right)^{-1},
\label{ratesD2comp}
\end{equation}
(ii) excitation swaps $|12\rangle \leftrightarrow |21\rangle$ with rates
\begin{widetext}
\begin{equation}
\Upsilon_{12}^{(kl)}\! =\frac{U}{|\hat{{\bf r}}_k - \hat{{\bf r}}_l|^{12}}\left(\! 1\! +\!\left\{ \frac{1}{4}\! \sum_{m\neq k,l}\! \left[\frac{\left(R_c^6\! -\! R^6\right)\! n_{1}^{(m)}\! +\! \left(R^6\! -\! R_c^6\right)\! n_{2}^{(m)}}{|\hat{{\bf r}}_k - \hat{{\bf r}}_m|^6} +\frac{\left(R_c^6\! -\! R^6\right)\! n_{2}^{(m)}\! +\! \left(R^6\! -\! R_c^6\right)\! n_{1}^{(m)}}{|\hat{{\bf r}}_l - \hat{{\bf r}}_m|^6}\right] \right\}^2\right)^{-1}.
\label{ratesDfinal}
\end{equation}
\end{widetext}
[which is the form adopted by (\ref{trans2}) after introducing the simplifications of this and the previous section] and (iii) spontaneous decay from $|1\rangle$ or $|2\rangle$ to $|0\rangle$, with (constant) rate $\kappa$. Similarly derived effective dynamics involving a single Rydberg level (thus excluding excitation swaps) were recently studied experimentally, yielding an excellent agreement with theoretical predictions \cite{urvoy2015,valado2016,gutierrez2017}.

According to Eq.~(\ref{ratesD2comp}), transitions between the ground state and each excited state -- type (i) processes -- slow down in the vicinity of previously excited atoms. This excitation blockade leads to dynamic bottlenecks \cite{lesanovsky2013, valado2016, gutierrez2016}, analogous to kinetic constraints in glassy systems \cite{ritort2003, chandler2010}. Following the rationale behind the swap Monte Carlo studies discussed in the Introduction, we expect that type (ii) processes may unblock otherwise highly constrained configurations, as illustrated in Fig.~\ref{figswaps} (a). In the following section we show compelling numerical results that confirm these expectations.

\section{Impact of excitation swaps on dynamics} 

\subsection{Accelerating effect of swaps}

To gain insight into the role of excitation swaps, we study a chain of $N$ atoms with periodic boundary conditions via continuous-time Monte Carlo (CTMC) \cite{bortz1975,newman1999}. Atoms $1\leq k \leq N/2$ are initially in state $|1\rangle$ while atoms $N/2 < k \leq N$ are initially in state $|2\rangle$. For intra-level interactions appreciably stronger than the inter-level interactions, i.e. if $R$ is considerably larger than $R_c$, such an initial configuration is highly arrested, see Eq.~ (\ref{ratesD2comp}). This means that, in the absence of exchange processes ($U = 0$), it takes a long time for the system to `forget' its initial configuration, see Fig.~\ref{figswaps} (b). When exchange interactions are included (for swap strength $U>0$), the relaxation time is drastically reduced --- see a representative trajectory for $U=1$ in Fig.~\ref{figswaps} (c). Swaps start being effective at the domain boundaries, due to the proximity of different types of excitations [notice the power-law decay of the swap rates with distance in (\ref{ratesDfinal})]. Then they propagate through the chain, eventually erasing all trace of the initial configuration. By contrast, trajectories of the same duration with $U=0$ hardly show any departure from the initial configuration.

To quantify the acceleration caused by swaps, we use the persistence function $P(t)$, defined as the fraction of atoms that have not changed state since $t=0$. It is displayed in Fig.~\ref{figpers} (a) for several values of the intra-level interaction parameter $R>1$ and the swap rate $U$ (the inter-level interaction parameter is $R_c = 1$, and for the moment decay is excluded, $\kappa = 0$). For $R=4$ the initial state is highly arrested, and swaps dramatically shorten the time that the system needs to move away from it. This effect is attenuated when the dynamical arrest is reduced, as in the curve for $R=1.5$, which is only moderately larger than $R_c$. By inspecting these and many other persistence curves with different sets of parameters, we conclude that the accelerating role of swaps for a given swap strength $U$ is dramatic when the interaction parameters, $R$ and $R_c$, are sufficiently large so the dynamics is glassy \cite{perezespigares2018}, and one is considerably larger than the other. In fact, if the intra-level parameter $R$ and the inter-level parameter $R_c$ are very similar, an excitation swap is not more effective than a swap of particles with very similar diameters in a supercooled liquid: the swap would not make the local relaxation significantly easier, as the initial and final configuration would be similarly constrained. In this sense, the role that the interaction parameters $R$ and $R_c$ play in the swap mechanism is analogous to that of particle diameters in glassy liquids \cite{berthier2019}. The effect of swaps is enhanced in Rydberg gases with markedly different interaction parameters,  as in highly polydisperse atomistic mixtures \cite{berthier2016} or in coarse-grained glassy models with wide local softness distributions \cite{gutierrez2019}.

In Fig.~\ref{figpers} (a), for either value of the intra-level interaction parameter $R$, the decay of $P(t)$ is approximately logarithmic for large swap strength $U$. This dependence can by understood from the statistics of local persistence times $\tau$, i.e. the time it takes an atom to undergo a state change. The distribution of $\tau$ in logarithmic scale $\pi[\log_{10}(\tau)]$ is highly peaked for $U=0$ but becomes wider and flatter for larger $U$,  see  Fig.~\ref{figpers} (b). As the persistence is $P(t) = \int_t^\infty d\tau\, p[\tau] =  \int_t^\infty d\tau\, \tau^{-1}\, \pi[\log_{10}(\tau)]$, for $\log_{10}(\tau)$ uniformly distributed across a $\tau$ interval, we obtain $P(t) \sim - \log_{10}(t)$ for $t$ within that interval [for shorter (longer) times $P(t) = 1$ ($0$)]. This suggests that the logarithmic decay is connected to the broad distribution of local persistence times caused by excitation swaps.

Further information on the relaxation dynamics can be obtained from the global persistence time $t_p$, defined as the time it takes for the persistence function $P(t)$ to reach zero, which corresponds to the longest local persistence time $\tau$ in the system. These are shown in Fig.~\ref{figpers} (c) using the same color code as in panel (a). A non-zero $U$ is seen to cause a departure from the persistence line without swaps, $U=0$, if a threshold value of the intra-level interactions $R$ is exceeded. This threshold value becomes smaller with increasing swap strength $U$, which means that if $R$ is sufficiently large, a very small swap strength $U$ has a strong effect, while a larger $U$ is needed to accelerate a less constrained dynamics. Above the threshold, swaps lead to shorter persistence times and a slower (almost linear) growth with $R$. Reductions of timescales of up to four orders of magnitude are observed, and larger reductions could be in principle achieved (although at a high computational cost) by making the dynamics more constrained or increasing the swap strength or both.

\subsection{Interplay of swap and decay processes} 

The presence of spontaneous decay, which annihilates excitations irrespective of the state of their neighbours, imposes a further timescale which, despite its non-collective nature, is relevant for the effectiveness of swaps. In fact, swaps can only be effective if they take place on timescales that are shorter than the typical decay time (otherwise, constrained configurations automatically relax by the effect of decay). This is illustrated in Fig.~\ref{figpers} (c), where persistence times $t_p$ for a decay rate $\kappa = 10^{-6}$ are shown in gray, and can be compared to the persistence times in a situation without decay ($\kappa = 0$). When the typical decay time $\kappa^{-1}$ is longer than $t_p$ for $\kappa=0$, the persistence time is not affected by decay processes. This is why the role of decay in the figure (where $\kappa = 10^{-6}$) is negligible for vanishing swap strength $U=0$ when the intra-level interaction parameter satisfies $R \lessapprox 3$, and also for $U=0.01$ when $R  \lessapprox 4$ and for $U=1$ throughout the range of $R$ that has been considered. If, as $R$ is increased, the persistence time in the absence of decay starts to exceed $\kappa^{-1}$, then the actual $t_p$ becomes independent of $R$ (and essentially equal to $\kappa^{-1}$) when the decay is present [gray symbols in Fig.~\ref{figpers} (c)]. From this we conclude that excitation swaps accelerate the dynamics as long as the typical decay time is longer than the reduced persistence times caused by the swaps. Otherwise the relaxation does not depend on collective effects as it is driven by spontaneous decay.

\section{Irreversibility and entropy production} 

\subsection{Irreversible dynamics in the presence of spontaneous decay}

When the decay rate $\kappa$ is zero, the (uncorrelated) stationary state is such that each atom can be in any energy level $|0\rangle,\, |1\rangle,\, |2\rangle$ with the same probability \cite{gutierrez2016}. As the rate between any two configurations connected by a transition is the same in either direction, see Eqs.~(\ref{ratesD2comp}) and (\ref{ratesDfinal}), the dynamics trivially satisfies detailed balance, and therefore is reversible \cite{kelly2011}. This is expected to change for non-vanishing decay ($\kappa > 0$), as then the excitation of atom $k$ from the ground state $|0\rangle$ to e.g. state $|1\rangle$ occurs with a rate $\Gamma_{1}^{(k)}$ [Eq.~(\ref{ratesD2comp})], while a larger rate $\Gamma_{1}^{(k)} + \kappa$ is associated with a de-excitation $|1\rangle$ to $|0\rangle$, and the probabilities of the many-body configurations are different.

\begin{figure}[t]
\includegraphics[scale=0.19]{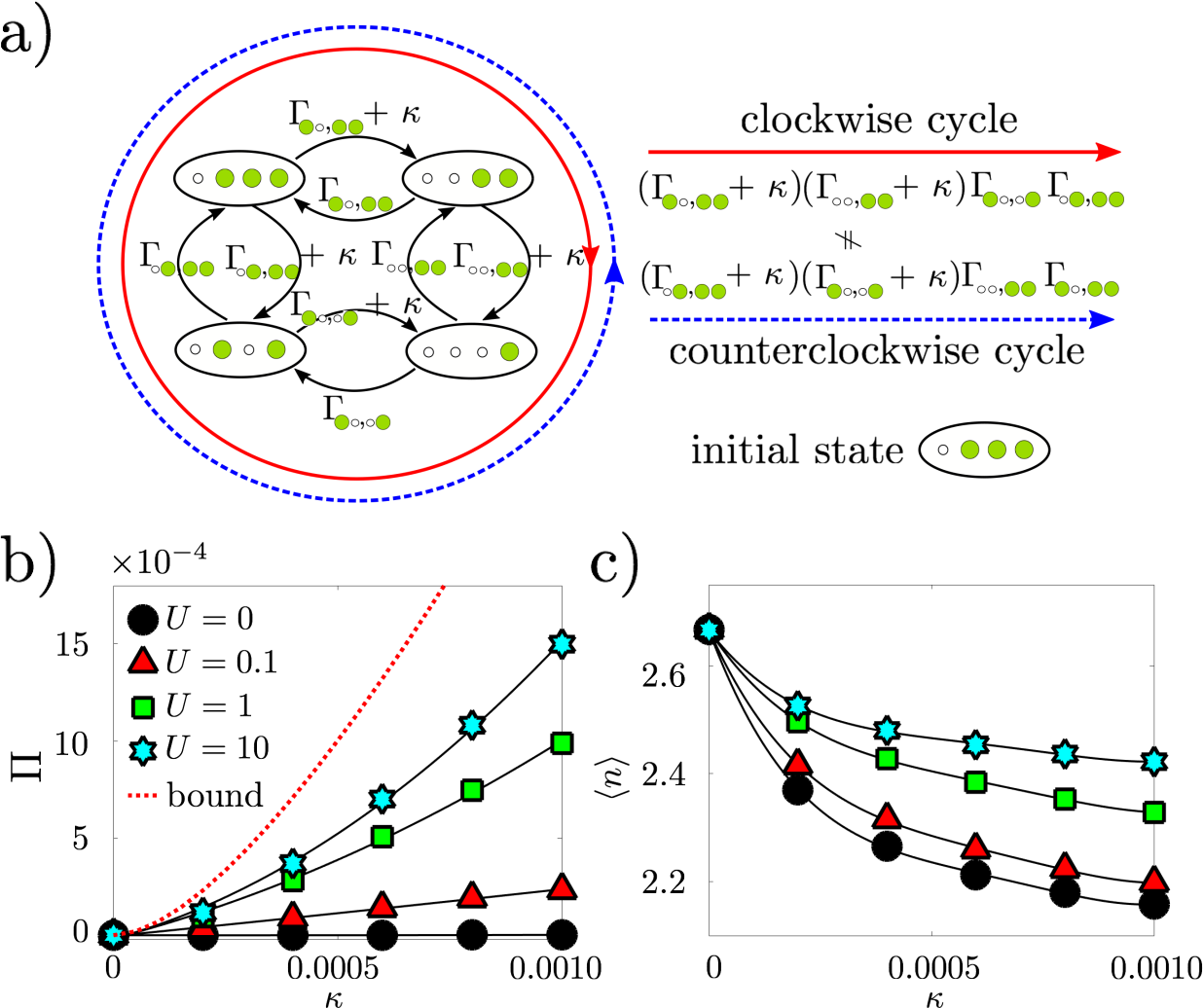}
\vspace{-0.2cm}
\caption{{\sf\bf Irreversibility and entropy production.}
(a) Cycle in configuration space. Subindices in the rates indicate the state of the neighbors of the atom undergoing the transition. The product of the transition rates depends on whether the cycle is traversed in a clockwise or a counterclockwise direction. The configuration on the top left corner has been chosen as the initial state. (b) Entropy production in a ring of $N=4$ atoms for $R=2$ and $R_c = 1$ as a function of the decay rate $\kappa$. Different swap strengths $U = 0, 0.1, 1, 10$ are considered (see legend). The entropy production bound in the limit of large $U$ [see Eq.~(\ref{bound})] is also included (red dotted line). (c) Average number of excitations in the stationary state $\langle n\rangle$ for the same parameter values as in (b).}\label{figentropy}
\end{figure}

To investigate the violation of detailed balance in the presence of decay we employ the Kolmogorov criterion. It states that a stationary Markov process is reversible if and only if the product of its transition rates along any cycle in configuration space is the same whether the path is traced in one or the other direction \cite{zia2007, kelly2011}. In Fig.~\ref{figentropy} (a) we show a simple cycle in configuration space along which two atoms are successively de-excited and excited back to their initial state. The pictorial notation for the excitation rates depicts the states of the first and second nearest neighbors to either side of the atom undergoing a transition (atoms immediately to the left and right of those shown are assumed to be excited). While the Kolmogorov criterion is seen to be satisfied for $\kappa=0$, in the presence of spontaneous decay, $\kappa > 0$, the product of the transition rates depends on the direction taken along the cycle, see Fig.~\ref{figentropy} (a). Thus, detailed balance is violated for $\kappa > 0$: the dynamics becomes irreversible due to the presence of decay processes.

\subsection{Entropy production rate}

The degree of irreversibility of a stochastic dynamics is captured by the entropy production rate \cite{schnakenberg1976,zia2007},
\begin{equation}
\Pi = \frac{1}{2} \sum_{\mu,\nu} \left(\Gamma_{\mu \to \nu}\, p_\mu - \Gamma_{\nu \to \mu}\, p_\nu\right) \log \frac{\Gamma_{\mu \to \nu}\, p_\mu}{\Gamma_{\nu \to \mu}\, p_\nu}.
\label{entprod}
\end{equation}
Here $\mu$ and $\nu$ are two configurations  (e.g. $|0 1 2 1 \cdots \rangle$ and $|1 1 2 1 \cdots \rangle$), $\Gamma_{\mu\to\nu}$ is the rate of the transition $\mu \to \nu$, and $p_\mu$ is the probability of configuration $\mu$ in the stationary state. The entropy production rate $\Pi$ is zero for reversible dynamics and positive otherwise. Eq.~(\ref{entprod}) can be simplified by removing the probabilities $p_\mu$ from the logarithm, as they account for the internal entropy production rate,
which vanishes in the stationary state  \cite{schnakenberg1976}. This yields a somewhat simpler expression: $\Pi = \frac{1}{2} \sum_{\mu,\nu} \left(\Gamma_{\mu \to \nu}\, p_\mu - \Gamma_{\nu \to \mu}\, p_\nu\right) \log(\Gamma_{\mu \to \nu}/\Gamma_{\nu \to \mu})$. The entropy production rate is accessible experimentally by continuously monitoring the states of each atom in a small cold-atomic lattice system or tweezer array \cite{schauss2012,saffman2016,labuhn2016,jau2016,bernien2017,kim2018,shi2018}, and inferring the probabilities $p_\mu$, and also the rates $\Gamma_{\mu \to \nu}$ via the so-called empirical currents, which count the number of transitions in a given time window.

The transition rates $\Gamma_{\mu \to \nu}$ in this context are: (A) $\Gamma_{1,2}^{(k)}$ if $\nu$ has an excited atom that is in the ground state in $\mu$, (B)  $\Gamma_{1,2}^{(k)}  + \kappa$ if $\nu$ has a ground state atom that is excited in $\mu$, and (C) $\Upsilon_{12}^{(kl)}$ if $\mu$ and $\nu$ are connected by an excitation swap. Case (C) does not contribute directly to the entropy production rate, as $\log(\Upsilon_{12}^{(kl)}/\Upsilon_{21}^{(kl)}) = 0$  ($\Upsilon_{12}^{(kl)}=\Upsilon_{21}^{(kl)}$). But excitation swaps do play a crucial role in a different way, as they modify the stationary distribution $p_\mu$ (see below).  Eq.~(\ref{entprod})  becomes
\begin{widetext}
\begin{equation}
\Pi\! =\! \sum_{\mu}\! \sum_{k_0}\! \sum_{e=1,2}\!\left( \kappa\ p_{\mu e^{(k_0)}}(U,\kappa)\! -\! \Gamma_{e}^{(k_0)}[p_\mu(U,\kappa)\!-\!p_{\mu e^{(k_0)}}(U,\kappa)] \right)\! \log \frac{\Gamma_{e}^{(k_0)}\!+\!\kappa}{\Gamma_{e}^{(k_0)}}.
\label{entprod2}
\end{equation}
\end{widetext}
The terms in brackets have been rearranged slightly for convenience. Here, $\mu$ is a given configuration and $\mu e^{(k_0)}$ is a configuration that is reached from $\mu$ by the excitation of site $k_0$ to level $|e\rangle$. (Strictly speaking, $\mu$ must have at least one atom in the ground state $|0\rangle$ or the corresponding term will be zero.) This transition occurs at a rate which we denote $\Gamma_{e}^{(k_0)}$. As the stationary probability distribution depends on the values of $U$ and $\kappa$, we explicitly write this dependence in $p_\mu(U,\kappa)$ and $p_{\mu e^{(k_0)}}(U,\kappa)$. In the sums, $k_0$ goes over all atoms in the ground state in $\mu$, and $e=1,2$ denotes excitations to state $|1\rangle$ or $|2\rangle$. For $\kappa = 0$, the logarithm vanishes, and so does $\Pi$; otherwise the logarithm is positive, and $p_\mu(U,\kappa) \geq p_{\mu e^{(k_0)}}(U,\kappa)$.

\subsection{The role of swaps in the entropy production}

In Fig.~\ref{figentropy} (b), we show the entropy production as a function of the decay rate $\kappa$ for different values of the swap strength $U$. To compute the entropy production, we obtain numerically the stationary distribution $p_\mu$ by running a CTMC simulation for a sufficiently long time so that the system has visited every configuration at least $10^4$ times. Then we simply apply Eq.~(\ref{entprod2}), as the rates for each $\mu$ are defined in Eq.~(\ref{ratesD2comp}). As estimating $p_\mu$ for each configuration $\mu$ is computationally demanding, we base our results on a small ring of $N=4$ atoms. 

As expected, the entropy production $\Pi$ is zero for $\kappa = 0$ (reversible dynamics) and increases for larger $\kappa$. More interestingly, for $U > 0$ the existence of excitation swaps leads to a significant increase in $\Pi$, which becomes more pronounced as $U$ gets larger. Somewhat counterintuitively, this is connected to the appearance of ``reversible'' shortcuts [in the sense that their associated probability currents, $\Gamma_{\mu \to \nu}\, p_\mu - \Gamma_{\nu \to \mu}\, p_\nu$, vanish  \cite{zia2007}, and they do not contribute directly to the entropy production, as pointed out above]. These shortcuts move the system between configurations with several excitations that are not connected by a single transition when swaps are absent.

While swaps by themselves do not produce entropy, they open alternative routes to highly inaccessible states. For example, a state with many excitations may be practically unreachable from the highly constrained states that have just one excitation less, given the very small rates $\Gamma_{1,2}^{(k)}$, but it may be reached through an appropriate sequence of excitation swaps. Exchange processes thus lead to a stationary distribution where on average the populations $p_{\mu e^{(k_0)}}(U,\kappa)$ are larger and closer to their associated $p_\mu(U,\kappa)$. The basic mechanism is illustrated by a simple model in Appendix D. This means that, as the swap strength $U$ grows, the positive term in brackets in Eq.~(\ref{entprod2}) becomes larger and the absolute value of the negative term becomes smaller. As the stationary probabilities are the only variables in Eq.~ (\ref{entprod2}) that are affected by swaps, the entropy production thus grows with $U$. To give numerical support to this observation, we show in Fig.~\ref{figentropy} (c) the average number of excitations in the stationary state $\langle n \rangle = \sum_\mu p_\mu(U,\kappa)\, n_\mu$ as a function of the decay rate $\kappa$ for different values of the swap strength $U$. Here the sum goes over all configurations and $n_\mu$ gives the number of sites that are in states $|1\rangle$ or $|2\rangle$ in $\mu$. Spontaneous decay suppresses the probability of configurations with a large number of excitations, but swaps counteract this effect to a significant extent
\footnote{The value that the average number of excitations
$\left< n \right>$
takes for $\kappa = 0$ in Fig.~\ref{figentropy} (c) is the value corresponding to a uniform stationary distribution. This is $\sum_{n=0}^N n\, {N \choose n}\, 2^n /3^N = 2N/3$ in a system of $N$ atoms, as can be easily seen from the binomial identity $x \frac{d}{dx} (1 + x)^N =   \sum_{n=0}^N n\, {N \choose n}\, x^n$.}.

To conclude, we derive an upper bound for the entropy production for large swap strength $U$. In that limit, $p_\mu \approx p_{\mu e^{(k_0)}} \approx 1/3^N$, i.e. the probability distribution approaches that of a completely mixed (infinite temperature) stationary state, and the term in brackets in Eq.~ (\ref{entprod2}) reaches its maximum, $\kappa/3^N$.  The entropy production then approaches
\begin{equation}
\Pi_{U\to\infty} = \frac{2 N \kappa}{3}\ \overline{\log (1 + \kappa/\Gamma_{e}^{(k_0)})},
\label{bound}
\end{equation}
which includes the average value of $\log (1 + \kappa/\Gamma_{e}^{(k_0)})$ taken over all $2 N 3^{N-1}$ excitation processes. This bound increases with the decay rate $\kappa$ and can be easily computed for small systems [see the red dotted line in Fig.~\ref{figentropy} (b)].

\section{Outlook}  

Excitation swaps dramatically shorten the relaxation timescales of dissipative Rydberg gases. This is what particle swaps in glassy soft-matter models also do, but there is one crucial difference: while swaps in those systems are just a numerical trick to reduce computational times, in atomic ensembles they are part of the physical dynamics. The accelerating effect of swaps in Rydberg gases is robust and does not require any fine-tuning of parameters, and therefore should be observable in modern cold atomic settings.  In fact, the observation of this effect only requires sufficiently strong interactions (which give rise to relaxation timescales much longer than the excitation timescale of isolated atoms), and sufficiently different interaction strengths for different excited levels. (Analogously, particle swaps are effective in liquids with glassy dynamics and if the model includes particles with sufficiently different diameters.) Additionally, there is the somewhat trivial requirement that decay processes must not take place predominantly on timescales that are shorter than the effect of swaps.

We have also shown that, apart from accelerating the dynamics, excitation swaps increase the entropy production rate, effectively moving the system further away from equilibrium conditions. This increased irreversibility is due to a reconfiguration of the stationary state distribution caused by the appearance of shortcuts in configuration space that link highly constrained configurations. The entropy production rate is in principle accessible in Rydberg quantum simulators and may enable a new handle for the characterization of non-equilibrium states in open quantum systems.

\begin{acknowledgments}
We are grateful to M. Marcuzzi and P. Rotondo for insightful discussions. The research leading to these results has received funding from the European Research Council under the European Union's Seventh Framework Programme (FP/2007-2013) / ERC Grant Agreement No. 335266 (ESCQUMA) and the EPSRC Grants No. EP/M014266/1, EP/R04421X/1, EP/R04340X/1. RG acknowledges the funding received from the European Union's Horizon 2020 research and innovation programme under the Marie Sklodowska-Curie Grant Agreement No. 703683. IL gratefully acknowledges funding through the Royal Society Wolfson Research Merit Award. We are also grateful for access to the University of Nottingham High Performance Computing Facility, and for the computing resources and related technical support provided by CRESCO/ENEAGRID High Performance Computing infrastructure (funded by ENEA, the Italian National  Agency for New Technologies, Energy and Sustainable Economic Development and 
by Italian and European research programmes) and its staff \cite{ponti2014}.
\end{acknowledgments}


\newpage

\appendix

\section{Derivation of Eq.~(\ref{integrand2})}

The integral in Eq.~(\ref{effectivedynpre}) contains all the (de-)excitation processes as well as the excitation swaps in the effective dynamics. The decay processes appear in the first term  $\mathcal{P}\mathcal{L}_1\mu$, but not in the integral, as the action of the corresponding dissipator on a matrix previously acted upon with $\mathcal{P}$ remains in the diagonal, and vanishes when projected under $\mathcal{Q}$. Consequently, the integrand in (\ref{effectivedynpre}) is
\begin{widetext}
\begin{eqnarray}
&\mathcal{P} \mathcal{L}_1 \mathcal{Q} e^{\mathcal{L}_0 t} \mathcal{Q} \mathcal{L}_1 \mu = &- \mathcal{P}\! \left( \sum_{s s^\prime}  \sum_{kl}  \Omega_{s^\prime} \Omega_s [\sigma_{s^\prime x}^{(l)}\,, e^{\mathcal{L}_0 t}  [\sigma_{s x}^{(k)}, \mu]]\right)\! -\! \mathcal{P} \left(\!\sum_{s<s^\prime} \sum_{s^{\prime\prime}} \sum_{k < l}  \sum_{m} E_{kl}^{ss^\prime}  \Omega_{s^{\prime\prime}} [\sigma_{s^{\prime\prime} x}^{(m)}\,, e^{\mathcal{L}_0 t}  [ \sigma_{ss^\prime}^{(k)} \sigma_{s^\prime s}^{(l)}+ \sigma_{s^\prime s}^{(k)} \sigma_{s s^\prime}^{(l)}, \mu]]\!\right) \nonumber\\
& & - \mathcal{P}\! \left(\sum_{s^\prime<s^{\prime\prime}} \sum_{s} \sum_{l < m}  \sum_{k} E_{lm}^{s^\prime s^{\prime\prime}}  \Omega_{s} [\sigma_{s^\prime s^{\prime\prime}}^{(l)} \sigma_{s^{\prime \prime} s^\prime}^{(m)}+ \sigma_{s^{\prime \prime} s^\prime}^{(l)} \sigma_{s^\prime s^{\prime\prime}}^{(m)}\,, e^{\mathcal{L}_0 t}  [\sigma_{s x}^{(k)} , \mu]]\right)\nonumber\\
& & - \mathcal{P}\! \left(\sum_{s<s^\prime} \sum_{s^{\prime\prime}<s^{\prime\prime\prime}}  \sum_{k < l}  \sum_{m < n} E_{kl}^{ss^\prime}  E_{mn}^{s^{\prime\prime}s^{\prime\prime\prime}} [\sigma_{s^{\prime\prime}s^{\prime\prime\prime}}^{(m)} \sigma_{s^{\prime\prime\prime}s^{\prime\prime}}^{(n)}\!+\! \sigma_{s^{\prime\prime\prime}s^{\prime\prime}}^{(m)} \sigma_{s^{\prime\prime}s^{\prime\prime\prime}}^{(n)}\,, e^{\mathcal{L}_0 t}  [\sigma_{ss^\prime}^{(k)} \sigma_{s^\prime s}^{(l)}\!+\! \sigma_{s^\prime s}^{(k)} \sigma_{s s^\prime}^{(l)} , \mu]]\!\right)
\label{integrand}
\end{eqnarray}
\end{widetext}
The second and third terms on the right hand side of the equation contain operators such as $\sigma_{ss^\prime}^{(k)} \sigma_{s^\prime s}^{(l)}$ that move off the diagonal the terms of $\mu$ corresponding to two different sites. Those operators are preceded or followed by operators such as $\sigma_{s^\prime x}^{(k)}$ that act on just one site, so the final result vanishes when acted upon by $\mathcal{P}$. A similar reasoning helps to simplify the two remaining terms. The first term simplifies because the action of $\sigma_{sx}^{(k)} = \left|s\right>_k\!\left<0\right| + \left|0\right>_k\!\left<s\right|$ combined with that of $\sigma_{s^\prime x}^{(m)} = \left|s^\prime\right>_m\!\left<0\right| + \left|0\right>_m\!\left<s^\prime\right|$ can only produce non-zero diagonal elements if $s^\prime = s$ and $m = k$. Analogously, in the fourth term only summands for which $k=m$, $l=n$, $s=s^{\prime\prime}$ and $s^\prime=s^{\prime\prime\prime}$ can yield nonzero contributions. Taking these facts into account simplifies the expression in Eq.~(\ref{integrand}) to that in Eq.~(\ref{integrand2}).

\section{Driving term $D[\mu,t]$}

The driving term $D[\mu,t]$ in Eq.~(\ref{effectivedynpre2}) is
\begin{eqnarray}
D[\mu,t]\!&=&\! -\! \sum_{s}\! \sum_{k} \Omega_s^2  \, \mathcal{P}\! \left( \sigma_{sx}^{(k)} e^{\mathcal{L}_0 t}\! \left(\sigma_{sx}^{(k)} \mu\right) -\sigma_{sx}^{(k)} e^{\mathcal{L}_0 t}\! \left(\mu\, \sigma_{sx}^{(k)}\right)\right.\nonumber\\
&-& \left.e^{\mathcal{L}_0 t} \left(\sigma_{sx}^{(k)} \mu\right) \sigma_{sx}^{(k)} +  e^{\mathcal{L}_0 t}  \left(\mu\, \sigma_{sx}^{(k)}\right) \sigma_{sx}^{(k)}\right).
\label{integrand_a}
\end{eqnarray}
For concreteness, we focus on the contribution of level $\left|1\right>$ on the right hand side. The first term yields
\begin{widetext}
\begin{eqnarray}
&\sigma_{1x}^{(k)} e^{\mathcal{L}_0 t} \left(\sigma_{1x}^{(k)} \mu\right) &= \sigma_{1x}^{(k)} e^{\mathcal{L}_0 t}  \left( \begin{array}{cccc}
0  &  \cdots & 0 & 0 \\
\vdots  &  \ddots & \vdots & \vdots \\
0 & \cdots & 0 & \rho_{00}^{(k)} \\
0 & \cdots & \rho_{11}^{(k)} &  0  \end{array} \right) 
= \sigma_{1x}^{(k)} e^{-i H_0 t}  \left( \begin{array}{cccc}
0  &  \cdots & 0 &  0 \\
\vdots  &  \ddots & \vdots & \vdots \\
0  &  \cdots & 0 &  e^{-\frac{1}{2} \gamma_1 t}\rho_{00}^{(k)} \\
0  &  \cdots  & e^{-\frac{1}{2} \gamma_1 t} \rho_{11}^{(k)} &  0  \end{array} \right) e^{i H_0 t}\\ \nonumber
& &= \left( \begin{array}{ccc}
\ddots & \vdots & \vdots \\
\cdots & e^{-\frac{1}{2} \gamma_1 t} e^{i t \sum_m \left[V^1_{km} n_1^{(m)} + \sum_s V^{1 s}_{km} n_s^{(m)}\right]} \rho_{11}^{(k)} &  0 \\
\cdots & 0 &  e^{-\frac{1}{2} \gamma_1 t} e^{-i t \sum_m \left[ V^1_{km} n_1^{(m)} +  \sum_s V^{1 s}_{km} n_s^{(m)}\right]} \rho_{00}^{(k)}  \end{array} \right)
\end{eqnarray}
\end{widetext}
The other terms can be similarly worked out. In the following, we use $\mathcal{V}^k_s = \sum_m \left[ V^s_{km} n_s^{(m)} + \sum_{s^\prime\neq s} V^{ss^\prime}_{km} n_{s^\prime}^{(m)}\right]$ as shorthand for the increment in the interaction energy corresponding to the excitation of atom $k$ to level $|s\rangle$. The term corresponding to $s=1$ in Eq.~(\ref{integrand_a}) becomes
\begin{widetext}
\begin{eqnarray} -\Omega_1^2  \left( \begin{array}{ccc}
\ddots & \vdots & \vdots \\
\cdots & 2\, e^{-\frac{1}{2} \gamma_1 t} \cos\left({\mathcal{V}^k_1 t}\right) \left[\rho_{11}^{(k)}-\rho_{00}^{(k)}\right]&  0 \\
\cdots & 0 &  2\, e^{-\frac{1}{2} \gamma_1 t} \cos\left(\mathcal{V}^k_1 t\right) \left[\rho_{00}^{(k)}-\rho_{11}^{(k)}\right] \end{array} \right),
\end{eqnarray}
\end{widetext}
and those contributions due to the other levels take an analogous form.  We thus obtain
\begin{equation}
D[\mu,t]\! =\!-\!\sum_{s=1}^p \Omega_s^2\, \sum_k\! 2 e^{-\frac{\gamma_s}{2} t} \cos\left({\mathcal{V}^k_s t}\right) \left[ \mathcal{I}_s^{(k)} \mu\! -\! \sigma_{sx}^{(k)} \mu\sigma_{sx}^{(k)}\right]\!,
\end{equation}
where the projection operator $\mathcal{I}_s^{(k)} = n_{s}^{(k)} + |0\rangle_k \langle 0|$. The effective dynamics due to the driving in $\mathcal{H}_1$ is given by
\begin{equation}
\int_0^\infty\! dt D[\mu,t]\! = \sum_{s=1}^p  \sum_k \frac{ 4 \Omega_s^2/\gamma_s}{1 + (2 \mathcal{V}_s^k/\gamma_s)^2}  \left[\sigma_{sx}^{(k)} \mu\sigma_{sx}^{(k)} -  \mathcal{I}_s^{(k)} \mu\right]\!.
\label{effectivedynpreintpspec}
\end{equation}

\section{Exchange term: $E[\mu,t]$}

The excitation swap term $E[\mu,t]$ in Eq.~(\ref{effectivedynpre2}) is 

\begin{eqnarray}
&&E[\mu,t] = - \sum_{s<s^\prime} \sum_{k < l}\left(E_{kl}^{ss^\prime}\right)^2 \mathcal{P} \left([\sigma_{s s^{\prime}}^{(k)} \sigma_{s^{\prime}s}^{(l)}\,, e^{\mathcal{L}_0 t}  [\sigma_{ss^\prime}^{(k)} \sigma_{s^\prime s}^{(l)}, \mu]]\right.\nonumber\\
&&+\!\left.[\sigma_{s s^{\prime}}^{(k)} \sigma_{s^{\prime}\!s}^{(l)}\,, e^{\mathcal{L}_0 t}  [\sigma_{s^\prime s}^{(k)} \sigma_{s s^\prime}^{(l)} , \mu]]\!+\![\sigma_{s^{\prime}\!s}^{(k)} \sigma_{s s^{\prime}}^{(l)}\,, e^{\mathcal{L}_0 t}  [\sigma_{ss^\prime}^{(k)} \sigma_{s^\prime\! s}^{(l)} , \mu]]\right.\nonumber\\
&&+\left.[\sigma_{s^{\prime}s}^{(k)} \sigma_{s s^{\prime}}^{(l)}\,, e^{\mathcal{L}_0 t}  [\sigma_{s^\prime s}^{(k)} \sigma_{s s^\prime}^{(l)} , \mu]]\right)\!.
\label{E}
\label{Eexplained}
\end{eqnarray}
Cancellations occur as diagonal terms are moved off and not brought back on the diagonal, which then vanish under the action of the projector $\mathcal{P}$, resulting in 
\begin{widetext}  
\begin{eqnarray}
E[\mu,t]&= &- \sum_{s<s^\prime} \sum_{k < l}\left(E_{kl}^{ss^\prime}\right)^2 \mathcal{P} \left( \sigma_{s s^{\prime}}^{(k)} \sigma_{s^{\prime}s}^{(l)}\, e^{\mathcal{L}_0 t} \left(\sigma_{s^\prime s}^{(k)} \sigma_{s s^\prime}^{(l)} \mu \right)- \sigma_{s s^{\prime}}^{(k)} \sigma_{s^{\prime}s}^{(l)}\, e^{\mathcal{L}_0 t} \left(\mu\, \sigma_{s^\prime s}^{(k)} \sigma_{s s^\prime}^{(l)} \right) -e^{\mathcal{L}_0 t} \left(\sigma_{s^\prime s}^{(k)} \sigma_{s s^\prime}^{(l)} \mu \right) \sigma_{s s^{\prime}}^{(k)} \sigma_{s^{\prime}s}^{(l)}\right.\nonumber\\
& &+  e^{\mathcal{L}_0 t} \left(\mu\, \sigma_{s^\prime s}^{(k)} \sigma_{s s^\prime}^{(l)} \right)  \sigma_{s s^{\prime}}^{(k)} \sigma_{s^{\prime}s}^{(l)} + \sigma_{s^{\prime}s}^{(k)} \sigma_{s s^{\prime}}^{(l)} e^{\mathcal{L}_0 t} \left(\sigma_{ss^\prime}^{(k)} \sigma_{s^\prime s}^{(l)} \mu\right)- \sigma_{s^{\prime}s}^{(k)} \sigma_{s s^{\prime}}^{(l)} e^{\mathcal{L}_0 t} \left(\mu\,\sigma_{ss^\prime}^{(k)} \sigma_{s^\prime s}^{(l)} \right)-e^{\mathcal{L}_0 t}\left(\sigma_{ss^\prime}^{(k)} \sigma_{s^\prime s}^{(l)} \mu\right)\sigma_{s^{\prime}s}^{(k)} \sigma_{s s^{\prime}}^{(l)}\nonumber\\
& &\left.+e^{\mathcal{L}_0 t}\left(\mu\,\sigma_{ss^\prime}^{(k)} \sigma_{s^\prime s}^{(l)} \right)\sigma_{s^{\prime}s}^{(k)} \sigma_{s s^{\prime}}^{(l)}\right).
\label{Esimplified}
\end{eqnarray}
\end{widetext}
Each term of the type $\sigma_{s s^{\prime}}^{(k)} \sigma_{s^{\prime}s}^{(l)}\ e^{\mathcal{L}_0 t} \left(\sigma_{s^\prime s}^{(k)} \sigma_{s s^\prime}^{(l)} \mu \right)$ can be unravelled as follows
\begin{eqnarray}
& &\sigma_{s s^{\prime}}^{(k)} \sigma_{s^{\prime}s}^{(l)}\ e^{\mathcal{L}_0 t} \left(\sigma_{s^\prime s}^{(k)} \sigma_{s s^\prime}^{(l)} \mu \right) =\sigma_{s s^{\prime}}^{(k)} \sigma_{s^{\prime}s}^{(l)}\ e^{\mathcal{L}_0 t} \left[\rho_{s s,s^\prime s^\prime}^{(k,l)} \right]_{s^\prime s, s s^\prime}\nonumber\\
& &= \sigma_{s s^{\prime}}^{(k)} \sigma_{s^{\prime}s}^{(l)}e^{-(\gamma_s+\gamma_s^\prime)t}  e^{-iH_0 t} \left[\rho_{s s,s^\prime s^\prime}^{(k,l)} \right]_{s^\prime s, s s^\prime}  e^{iH_0 t}\nonumber\\
& &=e^{-(\gamma_s+\gamma_s^\prime)t}  e^{-i[\mathcal{V}^k_{s^\prime}-\mathcal{V}^k_s + \mathcal{V}^l_s -  \mathcal{V}^l_{s^\prime}]t} \left[\rho_{s s,s^\prime s^\prime}^{(k,l)} \right]_{ss, s^\prime s^\prime}.
\label{termE1}
\end{eqnarray}
Here, $\left[\rho_{s^\prime s^\prime, s s}^{(k,l)} \right]_{ss^\prime,s^\prime s} $ denotes the $p^{N-2} \times p^{N-2}$ matrix defined by $\rho_{s^\prime s^\prime,ss}^{(k,l)} = \left({}_k{\left< s^\prime\right|}\otimes{}_l{\left< s\right|}\right) \rho \left(\left|s^\prime \right>_k \otimes\left| s\right>_l\right)$ located in the position of the (off-diagonal, and therefore rapidly decaying due to the dephasing) matrix element $\rho_{s s^\prime,s^\prime s}^{(k,l)}$, with all the other projected matrix elements being zero. We next work out the second term in Eq.~(\ref{Esimplified}):
\begin{eqnarray}
& &\sigma_{s s^{\prime}}^{(k)} \sigma_{s^{\prime}s}^{(l)}\, e^{\mathcal{L}_0 t} \left(\mu\, \sigma_{s^\prime s}^{(k)} \sigma_{s s^\prime}^{(l)} \right) =\sigma_{s s^{\prime}}^{(k)} \sigma_{s^{\prime}s}^{(l)}\ e^{\mathcal{L}_0 t} \left[\rho_{s^\prime s^\prime,s s}^{(k,l)} \right]_{s^\prime s, s s^\prime}\nonumber\\
&& = \sigma_{s s^{\prime}}^{(k)} \sigma_{s^{\prime}s}^{(l)}e^{-(\gamma_s+\gamma_s^\prime)t}  e^{-iH_0 t} \left[\rho_{s^\prime s^\prime,s s}^{(k,l)}  \right]_{s^\prime s, s s^\prime}  e^{iH_0 t}\nonumber\\
&&=e^{-(\gamma_s+\gamma_s^\prime)t}  e^{-i[\mathcal{V}^k_{s^\prime}-\mathcal{V}^k_s + \mathcal{V}^l_s -  \mathcal{V}^l_{s^\prime}]t} \left[\rho_{s^\prime s^\prime,s s}^{(k,l)} \right]_{ss, s^\prime s^\prime}. 
\label{termE2}
\end{eqnarray}
The third term  in Eq.~(\ref{Esimplified}) yields
\begin{eqnarray}
&&e^{\mathcal{L}_0 t} \left(\sigma_{s^\prime s}^{(k)} \sigma_{s s^\prime}^{(l)} \mu \right) \sigma_{s s^{\prime}}^{(k)} \sigma_{s^{\prime}s}^{(l)} = e^{\mathcal{L}_0 t} \left[\rho_{s s,s^\prime s^\prime}^{(k,l)} \right]_{s^\prime s, s s^\prime} \sigma_{s s^{\prime}}^{(k)} \sigma_{s^{\prime}s}^{(l)}\nonumber\\
&&= e^{-(\gamma_s+\gamma_s^\prime)t}  e^{-iH_0 t} \left[\rho_{s s,s^\prime s^\prime}^{(k,l)} \right]_{s^\prime s, s s^\prime}  e^{iH_0 t} \sigma_{s s^{\prime}}^{(k)} \sigma_{s^{\prime}s}^{(l)}\nonumber\\
& &=e^{-(\gamma_s+\gamma_s^\prime)t}  e^{-i[\mathcal{V}^k_{s^\prime}-\mathcal{V}^k_s + \mathcal{V}^l_s -  \mathcal{V}^l_{s^\prime}]t} \left[\rho_{s s,s^\prime s^\prime}^{(k,l)} \right]_{s^\prime s^\prime,ss}, 
\label{termE3}
\end{eqnarray}
and the fourth term yields
\begin{eqnarray}
&&e^{\mathcal{L}_0 t} \left(\mu\, \sigma_{s^\prime s}^{(k)} \sigma_{s s^\prime}^{(l)} \right)  \sigma_{s s^{\prime}}^{(k)} \sigma_{s^{\prime}s}^{(l)}= e^{\mathcal{L}_0 t} \left[\rho_{s^\prime s^\prime,s s}^{(k,l)} \right]_{s^\prime s, s s^\prime} \sigma_{s s^{\prime}}^{(k)} \sigma_{s^{\prime}s}^{(l)}\nonumber\\
&& = e^{-(\gamma_s+\gamma_s^\prime)t}  e^{-iH_0 t} \left[\rho_{s^\prime s^\prime,s s}^{(k,l)} \right]_{s^\prime s, s s^\prime}  e^{iH_0 t} \sigma_{s s^{\prime}}^{(k)} \sigma_{s^{\prime}s}^{(l)}\nonumber\\
& &=e^{-(\gamma_s+\gamma_s^\prime)t}  e^{-i[\mathcal{V}^k_{s^\prime}-\mathcal{V}^k_s + \mathcal{V}^l_s -  \mathcal{V}^l_{s^\prime}]t} \left[\rho_{s^\prime s^\prime,s s}^{(k,l)} \right]_{s^\prime s^\prime,ss}.
\label{termE4}
\end{eqnarray}
The remaining four terms  in (\ref{Esimplified}) can be easily worked out from (\ref{termE1},\ref{termE2},\ref{termE3},\ref{termE4}) by swapping the indices of the excited levels $s\leftrightarrow s^\prime$. Putting all these into Eq.~(\ref{Esimplified}),
\begin{widetext}
\begin{eqnarray}
E[\mu,t]&=&- \sum_{s<s^\prime} \sum_{k < l}\left(E_{kl}^{ss^\prime}\right)^2 e^{-(\gamma_s+\gamma_s^\prime)t} \left\{ e^{-i[\mathcal{V}^k_{s^\prime}-\mathcal{V}^k_s + \mathcal{V}^l_s -  \mathcal{V}^l_{s^\prime}]t}\,  \left(\left[\rho_{s s,s^\prime s^\prime}^{(k,l)}  - \rho_{s^\prime s^\prime,s s}^{(k,l)} \right]_{ss, s^\prime s^\prime} + \left[\rho_{s^\prime s^\prime,s s}^{(k,l)}-\rho_{s s,s^\prime s^\prime}^{(k,l)} \right]_{s^\prime s^\prime,ss}\right) \right. \nonumber\\
& &+ \left. e^{i[\mathcal{V}^k_{s^\prime}-\mathcal{V}^k_s + \mathcal{V}^l_s -  \mathcal{V}^l_{s^\prime}]t}\,  \left(\left[\rho_{s^\prime s^\prime,s s}^{(k,l)} - \rho_{s s,s^\prime s^\prime}^{(k,l)} \right]_{s^\prime s^\prime,ss} + \left[\rho_{s s,s^\prime s^\prime}^{(k,l)}- \rho_{s^\prime s^\prime, s s}^{(k,l)} \right]_{ss,s^\prime s^\prime}\right)\right\}\nonumber \\
&= &\! \sum_{s<s^\prime}\! \sum_{k < l}\!\left(E_{kl}^{ss^\prime}\right)^2\!e^{-(\gamma_s\!+\!\gamma_s^\prime)t}\, 2\cos{\left([\mathcal{V}^k_{s^\prime}\!-\!\mathcal{V}^k_s\! +\! \mathcal{V}^l_s\! -\!  \mathcal{V}^l_{s^\prime}]t\right)}\!  \left(\left[\rho_{s^\prime s^\prime,s s}^{(k,l)}\! -\! \rho_{s s,s^\prime s^\prime}^{(k,l)}\right]_{ss, s^\prime s^\prime}\! +\! \left[\rho_{s s,s^\prime s^\prime}^{(k,l)}\!-\!\rho_{s^\prime s^\prime,s s}^{(k,l)} \right]_{s^\prime s^\prime,ss}\right).\
\label{Esimplified2}
\end{eqnarray}
\end{widetext}
The exchange term  in Eq.~(\ref{effectivedynpre2}) can thus be written as
\begin{eqnarray}
& &\int_0^\infty\! dt\, E[\mu,t] = \sum_{s<s^\prime} \frac{2 ( \mathcal{E}_\beta^{ss^\prime})^2}{\gamma_s + \gamma_s^\prime} \sum_{k < l} \frac{1}{|{\bf r}_k - {\bf r}_l|^{2\beta}}\, \Gamma_{ss^\prime}^{(kl)}\times\nonumber\\
& &\left(\sigma_{s s^\prime}^{(k)} \sigma_{s^\prime s}^{(l)}\, \mu\, \sigma_{s^\prime s}^{(k)} \sigma_{s s^\prime}^{(l)} + \sigma_{s^\prime s}^{(k)} \sigma_{s s^\prime}^{(l)}\,  \mu\, \sigma_{s s^\prime}^{(k)} \sigma_{s^\prime s}^{(l)}\! -\!  \mathcal{I}_{ss^\prime}^{(k,l)} \mu\right),
\label{effectiveintpspecEB}
\end{eqnarray}
where the projector $\mathcal{I}_{ss^\prime}^{(k,l)} = n_s^{(k)} n_{s^\prime}^{(l)} +  n_{s^\prime}^{(k)} n_s^{(l)}$ and  $\Gamma_{ss^\prime}^{(kl)} = \left({1+\left[\frac{1}{\gamma_s\!+\!\gamma_s^\prime} \left(\mathcal{V}^k_{s^\prime}\!-\!\mathcal{V}^k_s\! +\! \mathcal{V}^l_s\! -\!  \mathcal{V}^l_{s^\prime}\right)\right]^2}\right)^{-1}\!.$

\section{Excitation swaps and redistribution of stationary-state probabilities}

\begin{figure}[t!]
\includegraphics[scale=0.29]{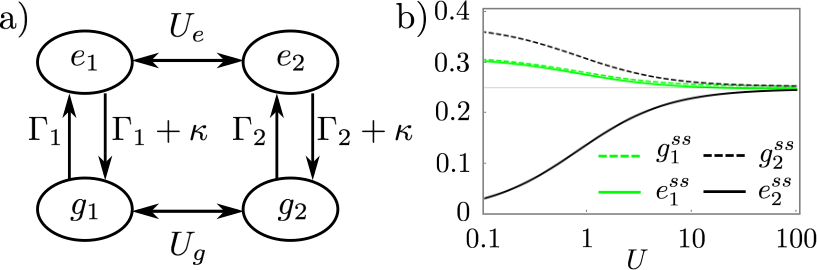}
\caption{ {\sf \bf Four-state model and stationary state probabilities as functions of the overall swap rate strength.} (a) Configurations, transitions and transition rates of the model. (b) Probabilities in the stationary state as a function of the overall swap strength $U$ for $U_g = 5\, U$ and $U_e = U$. The rates are $\Gamma_1 = 100$, $\Gamma_2 = 0.01$, and $\kappa = 1$, which satisfy $\Gamma_1 \gg \kappa \gg \Gamma_2$.}\label{figS1}
\end{figure}

We consider a simple model that illustrates how excitation swaps lead to a redistribution of probabilities in the stationary state. The configuration space comprises two ``ground'' states and two ``excited'' states, denoted as $|g_1\rangle$, $|g_2\rangle$, $|e_1\rangle$ and $|e_2\rangle$. From the perspective of the analogy between this abstract model and a Rydberg gas, the ``ground'' states may actually have excitations, as long as their number is smaller than the number of excitations in the corresponding ``excited'' states (we will not further consider the configuration of these states). The transitions between states are as follows [see Fig.~\ref{figS1} (a)].
\begin{itemize}
\item  The system can excite from $|g_1\rangle$ to $|e_1\rangle$ with a rate $\Gamma_1$, and from $|g_2\rangle$ to $|e_2\rangle$ with a rate $\Gamma_2$.
\item The system can de-excite from $|e_1\rangle$ to $|g_1\rangle$ with a rate $\Gamma_1+\kappa$, and from $|e_2\rangle$ to $|g_2\rangle$ with a rate $\Gamma_2+\kappa$. Here $\kappa$ plays the role of a decay rate.
\item  The system can undergo swap processes from $|e_1\rangle$ to $|e_2\rangle$ with a rate $U_e$, and from $|g_1\rangle$ to $|g_2\rangle$ with a rate $U_g$. The swap strength is smaller in the excited states $U_e \leq U_g$, but both are assumed to be proportional to an overall swap strength $U$.
\end{itemize}

The probabilities of each of the four states, which we denote as $g_1$, $g_2$, $e_1$ and $e_2$, evolve in time as follows
\begin{eqnarray}
\dot{g}_1 &= (\Gamma_1 + \kappa)\, e_1 + U_g\, g_2 - (\Gamma_1 + U_g)\, g_1 \label{g1}\\
\dot{g}_2 &= (\Gamma_2 + \kappa)\, e_2 + U_g\, g_1 - (\Gamma_2 + U_g)\, g_2 \label{g2}\\
\dot{e}_1 &= \Gamma_1\, g_1 + U_e\, e_2 - (\Gamma_1 + \kappa + U_e)\, e_1 \label{e1}\\
\dot{e}_2 &= \Gamma_2\, g_2 + U_e\, e_1 - (\Gamma_2 + \kappa + U_e)\, e_2. \label{e2} \nonumber
\end{eqnarray}
These equations are not independent, since they satisfy $\dot{g}_1 + \dot{g}_2 + \dot{e}_1 + \dot{e}_2 = 0$ by probability conservation. By setting three of them to zero, we obtain the stationary state values of e.g. $g_1$, $g_2$ and $e_1$ as functions of $e_2$. The normalization, $g_1 + g_2 + e_1 + e_2 = 1$, provides the remaining condition, yielding
\begin{widetext}
\begin{eqnarray}
g_1^{ss} &=\displaystyle \frac{U_e (\Gamma_1 + \kappa) \Gamma_2 + (\kappa^2 + \Gamma_1 \Gamma_2 + U_e (\Gamma_1 + \Gamma_2) + \kappa ( 2 U_e + \Gamma_1 + \Gamma_2)) U_g}{U_e (4 \Gamma_1 \Gamma_2 + \kappa(\Gamma_1 + \Gamma_2)) + (2 \kappa^2 + 4 \kappa U_e + 4 \Gamma_1 \Gamma_2 + 3 \kappa (\Gamma_1 + \Gamma_2) + 4 U_e (\Gamma_1 + \Gamma_2)) U_g} \label{g1ss}\\
g_2^{ss} &=\displaystyle \frac{U_e (\Gamma_2 + \kappa) \Gamma_1 + (\kappa^2 + \Gamma_1 \Gamma_2 + U_e (\Gamma_1 +\Gamma_2) + \kappa ( 2 U_e + \Gamma_1 + \Gamma_2)) U_g}{U_e (4 \Gamma_1 \Gamma_2 + \kappa(\Gamma_1 + \Gamma_2)) + (2 \kappa^2 + 4 \kappa U_e + 4 \Gamma_1 \Gamma_2 + 3 \kappa (\Gamma_1 + \Gamma_2) + 4 U_e (\Gamma_1 + \Gamma_2)) U_g} \label{g2ss}\\
e_1^{ss} &=\displaystyle\frac{U_e \Gamma_1 \Gamma_2 + \Gamma_1 (\Gamma_2 + \kappa) U_g + U_e (\Gamma_1 + \Gamma_2) U_g}{U_e (4 \Gamma_1 \Gamma_2 + \kappa(\Gamma_1 + \Gamma_2)) + (2 \kappa^2 + 4 \kappa U_e + 4 \Gamma_1 \Gamma_2 + 3 \kappa (\Gamma_1 + \Gamma_2) + 4 U_e (\Gamma_1 + \Gamma_2)) U_g} \label{e1ss}\\
e_2^{ss} &=\displaystyle \frac{U_e \Gamma_1 \Gamma_2 + \Gamma_2 (\Gamma_1 + \kappa) U_g + U_e (\Gamma_1 + \Gamma_2) U_g}{U_e (4 \Gamma_1 \Gamma_2 + \kappa(\Gamma_1 + \Gamma_2)) + (2 \kappa^2 + 4 \kappa U_e + 4 \Gamma_1 \Gamma_2 + 3 \kappa (\Gamma_1 + \Gamma_2) + 4 U_e (\Gamma_1 + \Gamma_2)) U_g} . \label{e2ss}
\end{eqnarray}
\end{widetext}
For $\kappa = 0$ we obtain a fully mixed state $g_1^{ss} = g_2^{ss} = e_1^{ss} =e_2^{ss} =\frac{1}{4}$ (and detailed balance is satisfied).

We now assume that $\Gamma_1 \gg \kappa \gg \Gamma_2$. This means that transitions between $|g_1\rangle$ and $|e_1\rangle$ occur almost as frequently in one direction as in the opposite direction ($\Gamma_1 + \kappa \approx \Gamma_1$), while excitations from $|g_2\rangle$ to $|e_2\rangle$ occur much less frequently than the opposite (de-excitation) process ($\Gamma_2 + \kappa \approx \kappa$). This assumption allows us to reproduce (in a simplified form) the dynamic heterogeneity of the full Rydberg dynamics, where both highly mobile ($|g_1\rangle$ and $|e_1\rangle$) and highly arrested ($|g_2\rangle$ and $|e_2\rangle$) configurations exist.

We first consider the situation when swap processes are absent, $U=0$. The stationary state probability distribution is then obtained by setting $U_e = U_g=0$ in Eqs.~(\ref{g1}, \ref{g2},\ref{e1},\ref{e2})  [or one can directly simplify Eqs.~(\ref{g1ss}, \ref{g2ss}, \ref{e1ss}, \ref{e2ss})], which yields 
\begin{equation}
e_1^{ss} = \frac{\Gamma_1}{\Gamma_1 + \kappa}g_1^{ss} \approx g_1^{ss},\ e_2^{ss} = \frac{\Gamma_2}{\Gamma_2 + \kappa}g_2^{ss} \ll g_2^{ss}.
\label{U0}
\end{equation}
These are two uncoupled two-level systems with very different stationary distributions: in one case the probabilities of ``ground'' state and ``excited'' state are almost identical, in the other the population of the ``ground'' state is much larger.

We then focus on the opposite situation, i.e. when swap processes occur very frequently. If both $U_e$ and $U_g$ are very large, the dominant contributions in Eqs.~(\ref{g1ss},  \ref{g2ss},  \ref{e1ss}, \ref{e2ss}) are those of second order in $U$, leading to 
\begin{eqnarray}
\lim_{U\to\infty} g_1^{ss} &=&\lim_{U\to\infty} g_2^{ss} =\displaystyle \frac{\Gamma_1 + \Gamma_2 + 2 \kappa}{4 (\kappa + \Gamma_1 + \Gamma_2)} \approx \frac{1}{4}; \label{largeU1}\\
\lim_{U\to\infty}e_1^{ss} &=& \lim_{U\to\infty}e_2^{ss} =\displaystyle\frac{\Gamma_1 + \Gamma_2}{4 (\kappa + \Gamma_1 + \Gamma_2)} \approx \frac{1}{4}. \label{largeU2}
\end{eqnarray}
We conclude that an enhancement of exchange processes eventually leads to a vanishing probability imbalance between ``ground'' and ``excited'' states, $e_2^{ss} \approx g_2^{ss}$. 

So far we have only considered the two limiting cases of vanishing swap strength or very large swap strength. In Fig.~\ref{figS1} (b), we show $g_1^{ss}$, $g_2^{ss}$, $e_1^{ss}$ and $e_2^{ss}$ as functions of $U$ for parameter values satisfying $\Gamma_1 \gg \kappa \gg \Gamma_2$ (similar results are obtained for other parameter choices that fulfill this condition). This illustrates the gradual change of the stationary state probabilities as the overall swap strength $U$ is increased. The curves smoothly join the solution in Eq.~(\ref{U0}) to that in Eqs.~(\ref{largeU1},\ref{largeU2}) as $U$ is increased. These results provide a simple picture of the basic mechanism underlying the much more complex situation considered in Section IV.C in general, and more specifically in the numerical results shown in Fig.~\ref{figentropy} (c).

\end{document}